\documentstyle[12pt]{article}
\makeatletter
\def\@maketitle{\newpage
 \null
 {\normalsize \tt \begin{flushright}
  \begin{tabular}[t]{l} \@date
  \end{tabular}
 \end{flushright}}
 \begin{center}
 \vskip 2em
 {\LARGE \@title \par} \vskip 1.5em {\large \lineskip .5em
 \begin{tabular}[t]{c}\@author
 \end{tabular}\par}
 \end{center}
 \par
 \vskip 1.5em}
\makeatother
\begin{document}
\setlength{\baselineskip}{16pt}
\title{Integrability of Schwinger-Dyson Equations \\
                     in 2D Quantum Gravity     and \\
                     $c < 1$ Non-critical String Field Theory}
\author{
        Ryuichi NAKAYAMA \thanks{
                          nakayama@particle.phys.hokudai.ac.jp}
      and Toshiya SUZUKI \thanks{
                          tsuzuki@particle.phys.hokudai.ac.jp}
\\[1cm]
{\small
    Department of Physics, Faculty of Science,} \\
{\small
           Hokkaido University, Sapporo 060, Japan}
}
%
\date{
  EPHOU-95002  \\
  HUPS-94-5     \\
  hep-th/9503190 \\
  March 1995
}


\maketitle

\begin{abstract}

We investigate the integrability of the Schwinger-Dyson equations in $c = 1 -
\frac{6}{m(m+1)}$ string field theory which were proposed by Ikehara et al as
the continuum limit of the Schwinger-Dyson equations of the matrix chain
model.  We show the continuum Schwinger-Dyson equations generate a closed
algebra.  This algebra contains Virasoro algebra but does not coincide with
$W_{\infty}$ algebra. We include in the Schwinger-Dyson equations a new
process of removing from the loop boundaries the operator ${\cal H}(\sigma)$
which locally changes the spin configuration.   We also derive the
string field Hamiltonian from the continuum Schwinger-Dyson equations.
Its form is universal for all $c = 1 - \frac{6}{m(m+1)}$ string theories.

\end{abstract}

\newpage

\newcommand {\n}{\nonumber\\}
\newcommand {\nn} {\nonumber}

\newcommand {\cleqn}{\setcounter{equation}{0}}
\renewcommand {\theequation}{\arabic{equation}}
\newcommand \eq[1]{(\ref{#1})}

\newcommand {\beq}{\begin{equation}}
\newcommand {\eeq}{\end{equation}}

\newcommand {\beqa}{\begin{eqnarray}}
\newcommand {\eeqa} {\end{eqnarray}}

\newcommand {\beqc}{\beq\begin{array}{c}}
\newcommand {\eeqc}[1]{\label{#1}\end{array}\eeq}

\newcommand {\beql}{\beq\begin{array}{l}}
\newcommand {\eeql}[1]{\label{#1}\end{array}\eeq}

\newcommand {\beqr}{\beq\begin{array}{r}}
\newcommand {\eeqr}[1]{\label{#1}\end{array}\eeq}

\newcommand {\lfrac}[2]{\frac{\displaystyle #1}{\displaystyle #2}}
\newcommand {\sfrac}[2]{\displaystyle \frac{#1}{#2}}

\newcommand {\dee}{\partial}
\newcommand {\deeb}{\bar{\partial}}

\newcommand \bra[1]{\left< {#1} \,\right\vert}
\newcommand \ket[1]{\left\vert\, {#1} \, \right>}
\newcommand \braket[2]{\hbox{$\left< {#1} \,\vrule\, {#2} \right>$}}

\newcommand {\lp}{\left(}
\newcommand {\rp}{\right)}

\newcommand {\ud}{\updownarrow}


String field theory \cite{kaku} seems to be the most natural framework for
studying the non-perturbative properties of string theory. Recently a new class
of string field theory  based on the transfer matrix formalism \cite{kkmw}
for 2d quantum gravity was proposed in \cite{ik}. In this string field
theory the geodesic distance from the boundaries is used as a time variable.
By using this time coordinate the world sheet of c = 0 string is cut into
time slices and then decomposed into vertices and propagators.  This string
field theory is called string field theory in the temporal
gauge\cite{fikn}\cite{naka}.

This decomposition is also possible even when matter degrees of freedom are
put on the worldsheet.  The simplest model of string with matter is the Ising
model on a random surface, {\it i.e.}, c = 1/2 string.  The partition
function and the loop amplitudes of the Ising model on a dynamically
triangulated surface \cite{kkm} are defined in terms of the two-matrix
model\cite{kaza} and the continuum theory is obtained by the double scaling
limit \cite{matrix}, which also enables us to discuss the summation of string
perturbation series. The most effective method for non-perturbative
investigation of the loop amplitudes will be  the Schwinger-Dyson equations
(SDEs). The SDEs in the matrix model determine the loop amplitudes completely.

The string field theory can be constructed
in such a way that the continuum version of the matrix model SDEs are derived
from the SDEs in string field theory.
In \cite{iikmns} the continuum limit of the SDEs in the two-matrix model was
derived under some assumptions and the string field Hamiltonian was inferred
from these equations.  Such assumptions were then justified by showing that
$W_3$ constraints \cite{fkn} \cite{dvv} \cite{gava} can be derived from the
continuum version of the matrix model SDEs. These results were also extended
to c = 1 $-$ 6/(m(m+1)) string.

The purpose of this letter is to investigate the integrability of the SDEs
proposed in \cite{iikmns}. The SDEs of c = 0 string are so-called Virasoro
constraints on the partition function \cite{fkn} \cite{dvv} and these SDEs
are integrable because Virasoro algebra closes. The SDEs for c = 1 $-$
6/(m(m+1)) string proposed in \cite{iikmns} are more complicated than those
for c = 0 string and the integrability of them is not obvious.
We will show these SDEs are indeed integrable by exhibiting the algebra
generated by these SDEs.  This will  give another justification for the
assumptions in \cite{iikmns}. Contrary to our expectations, however, this
algebra turns out to be not $W_{\infty}$ but a larger one. Furthermore this
algebra does not seem to contain $W_{\infty}$ as a subalgebra.
We will also include in the SDEs those terms corresponding  to a process of
removing the operator ${\cal H}(\sigma)$ from the boundary loops, which were
not taken into account in \cite{iikmns}.  This operator changes the
configuration of spins locally.\cite{iikmns}  We will then  write down the
Hamiltonian for c = 1 $-$ 6/(m(m+1)) string in a more explicit way than in
\cite{iikmns}, where the meaning of the summation over conformal field
theory (CFT) states $|v>$ in the Hamiltonian was not completely specified.

In \cite{ik2} another type of c = 1/2 string field theory was constructed by
changing the definition of the time coordinate in such a way that the string
does not cross the domain walls.  It was shown the SDEs generate decoupled
Virasoro algebras.  In \cite{mog} and \cite{kos} temporal-gauge string field
theory was extended to include open string fields in two different ways.
It was also pointed out in \cite{jr} that the string field Hamiltonian can
be derived from the stochastic quantization of the matrix model.  The string
field Hamiltonian was also deduced from the matrix model in \cite{wata}.

Let us consider a two-matrix model defined by an action
\beq
S(A,B) = N tr( \frac{1}{2} A^2 + \frac{1}{2} B^2 - c AB - \frac{1}{3} \lambda
(A^3 + B^3)) ,
\label{1}
\eeq
where $A$ and $B$ are N by N hermitian matrices and $c$ and $\lambda$
are constants.
This matrix model is known to provide  a formal perturbative definition of c
$=$ 1/2 non-critical string.\cite{kaza} Such a string can be realized by
putting
the Ising model on a dynamically triangulated surface.  When one assigns
the vertices $\lambda A^3$ and $\lambda B^3$ to triangles on which up
and down spins
reside, respectively, the free energy of the matrix model (\ref{1}) gives
the partition function  of the Ising model on a random surface.

The loop amplitudes in which all spins on the loop boundaries are up are
expressed by correlation functions of $W_0(m) = \frac{1}{N} tr A^m $.
Because the
action (\ref{1}) mixes $A$ and $B$, however, the
SDEs in the two-matrix model do not close within this kind of
loops but loops with mixed spin configurations appear.  Therefore we have to
consider correlation functions of the operator
\beq
W_n(m_1,m_2, \cdots ,m_n) = \frac{1}{N}  tr(A^{m_1} B A^{m_2} B \cdots
         A^{m_n} B)  \quad \qquad (n \geq 1).
\label{2}
\eeq
This operator represents a loop on which almost all spins are up and only
{\it n} non-adjacent spins are down.
By considering a correlation function of $(1/N)tr(t^a A^{m_1} B A^{m_2} B
\cdots B A^{m_s}) $ and $W_n$'s, where $ \{ t^a | a=1,\cdots,N^2 \} $ is
a basis of Lie algebra $u(N)$, and using the fact that this correlation
function is invariant under an infinitesimal shift $ \delta A = \epsilon t^a$
 of the integration variable $A$, we obtain an SDE for the loops (\ref{2}).
This SDE describes a process of taking away one triangle from a boundary and
contains terms which correspond to the following processes.
\begin{description}
\item[1.] The boundary loop splits into two.
\item[2.] The boundary loop absorbs another boundary.
\item[3.] The spin configuration on the boundary loop changes.
\end{description}
In \cite{iikmns} it was assumed that these terms survive the continuum
limit and the remaining terms   which change the length of the loop drop out.
With this assumption we can write down the continuum SDE. The continuum
limit of a loop is described by the length $l$ of the loop  and the state
of the  matter on the loop, which is a state in c = 1/2 conformal field
theory (CFT).  The continuum limit of the simplest loop operator
$W_0(m)=\frac{1}{N} tr A^m$ is specified by the
length $l$ and the  state $|+>$ in c = 1/2  CFT that corresponds
to the up spin. We will denote it by $\hat{w}_0 (l)$.
The continuum limit of (\ref{2}) is specified by the lengths
$\{ l_1, l_2, \cdots, l_n \}$ of the segments of $|+>$ state and will be
denoted by $\hat{w}_n (l_1, l_2, \cdots, l_n)$.
This can be obtained from $\hat{w}_0(l_1+\cdots l_n)$ by insertion of the
local operator ${\cal H}(\sigma)$, which flips the spin locally, at {\it n}
distinct points of the loop boundary.\cite{iikmns}
To write down the continuum SDE we introduce the source functions
$J_0 (l)$ for $\hat{w}_0 (l)$ and $J_n (l_1, l_2, \cdots, l_n)$ for
$\hat{w}_n(l_1, l_2, \cdots, l_n)$.  Then the SDEs in the continuum limit
should be
\beq
T_n (l_1, l_2, \cdots, l_n) Z[J]  \approx  0  \quad (n \geq 1),
\label{4}
\eeq
where $Z[J] = Z[J_0, J_1,J_2, \ldots]$ is the  generating functional of the
 disconnected loop amplitudes.  $T_n$ is a functional differential
operator with respect to $J$'s and given by
\beqa
T_1(l) & = & \int_0^l dl' D_0(l') D_0(l - l') + \int_0^{\infty} dl' J_0(l') l'
D_0(l+l')    \n
& & + \sum_{m=1}^{\infty} \sum_{j=1}^m \int_0^{\infty} dl_1' \cdots
\int_0^{\infty } dl_m' J_m(l_1',\cdots,l_m') l_j' D_m(l_1',\cdots,l_{j-1}',
l_j'+l,l_{j+1}',\cdots,l_m') \n
& & \quad \quad + D_1(l),
\label{5}
\eeqa
\beqa
T_n(l_1,\cdots,l_n) & = & \int_0^{l_1} dl' D_0(l')D_{n-1}
             (l_n+l_1-l',l_2,\cdots,l_{n-1}) \n
& & + \sum_{k=2}^{n-1} \int_0^{l_k} dl' D_{k-1}(l_1+l',l_2,\cdots,l_{k-1})
D_{n-k}(l_n+l_k-l',l_{k+1},\cdots,l_{n-1}) \n
& & + \int_0^{l_n} dl' D_{n-1}(l_1+l',l_2, \cdots,l_{n-1}) D_0(l_n-l') \n
& & + \int_0^{\infty} dl' J_0(l')l' D_{n-1}(l_n+l_1+l',l_2,\cdots,l_{n-1}) \n
& & + \sum_{m=1}^{\infty} \sum_{j=1}^{m} \int_0^{\infty} dl_1' \cdots
\int_0^{\infty} dl_m' J_m(l_1',\cdots,l_m') \int_0^{l_j'} dl'' \n
& & \quad \cdot D_{m+n-1} (l_1',\cdots,l_{j-1}',l_1+l'',l_2,\cdots,
l_{n-1},l_n+l_j'-l'',l_{j+1}',\cdots,l_m') \n
& & \quad \quad + D_n(l_1,\cdots,l_n), \quad \quad \quad (n \geq 2).
\label{6}
\eeqa
$D_0(l)$ and $D_n(l_1,\cdots,l_n)$ are the functional derivatives  defined by
 $D_0(l) J_0(l') = \delta(l - l')$ and
$D_n (1_1, \cdots, l_n) J_m (l'_1, \cdots, l'_m) = (1/n) \delta_{nm} (
\delta(1_1 - l'_1) \cdots \delta(l_n - l'_n)  + \mbox{ cyclic permutations}) $.
Here the string coupling constant $g$ is suppressed for simplicity.
These SDEs describe the above three processes.(figs 1 and 2)  The symbol
$\approx 0$ means
that the left  hand side is equal to a sum of terms proportional to
the  products of the delta
functions  $\delta(l)$ and $\delta(l_j)$ and their derivatives.
These terms represent the following two processes.
\begin{description}
\item[4.] When a triangle is taken away at the point where the spin is down,
the spin flips up.
\item[5.] A loop with vanishing length disappears.
\end{description}
{}From the point of view of string field theory  process 5 is expressed
by the tadpole terms.\cite{iikmns}  Process 4 is a new one that was not
taken into account in \cite{iikmns}. Later we will incorporate such a process
into (\ref{6}).

In \cite{iikmns} it was shown that a combination of a subset of SDE(\ref{4})
\beqa
T_1(l) Z[J] |_{(J_n = 0,  n \geq 1)} & \approx & 0,  \n
T_2(l,0) Z[J] |_{(J_n = 0,  n \geq 1)} & \approx & 0
\label{7}
\eeqa
and the \lq null' condition
\beq
D_2(l,0) Z[J] |_{(J_n = 0,  n \geq 1)} \approx 0
\label{8}
\eeq
is equivalent to the $W_3$ constraints \cite{fkn} \cite{dvv} \cite{gava}.
We can show that the commutator of $T_n(l_1,\cdots,l_n)$ and
$D_m(l_1',\cdots,l_{m-1}',0)$ is given by a linear combination of terms of the
form $D_{n+m-1}(
l_1'',\cdots,l_{n+m-2}'',0)$.  Hence the \lq null' condition $D_m(l_1,
\cdots,l_{m-1},0) Z \approx 0$ and the SDEs are compatible.

A central issue of this letter is the integrability of (\ref{4}).  SDEs in
c = 0 string can be succinctly written as Virasoro constraints \cite{fkn}
\cite{dvv} and the closure
of Virasoro algebra ensures the integrability of SDEs.  For c = 1/2 string
this algebra  is replaced by $W_3$ algebra \cite{fkn} \cite{dvv} \cite{gava}
and we might expect (\ref{5}) and (\ref{6}) generate $W_3$ algebra.
In \cite{ik2} an alternative definition of the time coordinate is chosen
in c = 1/2 string field theory and the resulting algebra is shown to
be decoupled Virasoro algebras.  We study whether $T_n$ generates a closed
algebra, and if it does, what the algebra is.
Here processes  4 and 5  will be ignored and later we will take only process
4 into account. The calculation is straightforward and we will
present only the results of long calculation.
\beq
[T_1(l),T_1(\tilde{l})]  =  (l - \tilde{l}) T_1(l + \tilde{l}),
\label{9}
\eeq
\beqa
\mbox{ [ } T_n(l_1, \cdots,l_n), T_1(l) \mbox{ ] } & = & \sum_{j = 1}^n l_j
T_n(l_1,\cdots,l_{j-1},l_j+l,l_{j+1},\cdots,l_n) \n
& & - \int_0^l dl' T_n(l_1+l',l_2,\cdots,l_{n-1},l_n+l-l') \n
& & \mbox{          } \quad \quad \quad \quad (n \geq 2),
\label{10}
\eeqa
\beqa
& & \mbox{ [ } T_n(l_1, \cdots,l_n), T_m(\tilde{l}_1,\cdots,\tilde{l}_m)
\mbox{ ] } \n
& = & \int_0^{l_1} dl' T_{n+m-1} (\tilde{l}_1 + l',\tilde{l}_2, \cdots,
\tilde{l}_{m-1},\tilde{l}_m+l_1-l',l_2,\cdots,l_n) \n
&  + & \sum_{j=2}^{n-1} \int_0^{l_j} dl' T_{n+m-1}(l_1,\cdots,l_{j-1},
\tilde{l}_1+l',\tilde{l}_2,\cdots,\tilde{l}_{m-1},\tilde{l}_m+l_j-l',l_{j+1},
\cdots,l_n) \n
& + &\int_0^{l_n} dl' T_{n+m-1} (l_1,\cdots,l_{n-1},\tilde{l}_1+l',
\tilde{l}_2,\cdots,\tilde{l}_{m-1},\tilde{l}_m+l_n-l') \n
& & \mbox{     } - (n \leftrightarrow m, l_j \leftrightarrow \tilde{l}_j)
        \quad \quad  (n, m \geq 2).
\label{11}
\eeqa
Hence  the algebra closes and SDEs (\ref{4}) are  integrable. The algebra
by itself is quite intriguing. While  (\ref{9}) is Virasoro algebra, the
whole algebra including the other two, (\ref{10}) and (\ref{11}), is a new one
 and there seems to be no  simple relationship to $W_3$ or $W_{\infty}$
algebra.

Let us next incorporate process 4.
This will be done by adding to $T_n$ some terms proportional to $D_{n-2}$
so that the algebra (\ref{9}) $\sim$ (\ref{11})
remains unchanged.  We replace $T_n (n \geq 2) $ by the following operator
$\tilde{T}_n$.
\beq
\tilde{T}_2(l_1,l_2) =   T_2(l_1,l_2)  +  \{ a+b( \dee_{l_1} + \dee_{l_2}) \}
 \{ ( \delta(l_1) + \delta(l_2) ) D_0(l_1 + l_2) \}, \label{11.5}
\eeq
\beqa
 \tilde{T}_n(l_1, \cdots,l_n) =  T_n(l_1,\cdots,l_n) + & & \n
   \{   a + b(\dee_{ l_1} + \dee_{ l_n}) \} \{ & \delta(l_1) &
            D_{n-2}(l_n+l_1+l_2,l_3,\cdots,l_{n-1}) \n
     + & \delta (l_n) & D_{n-2}(l_{n-1}+l_n+l_1,l_2 ,\cdots,l_{n-2}) \}  \n
     & & \quad (n \geq 3).
\label{12}
\eeqa
$T_1$ is left unchanged: $\tilde{T}_1(l) = T_1(l)$. The second terms in
(\ref{11.5}) and (\ref{12}) describe  process 4.(figs 3 and 4) Here $a$ and
$b$ are constants.
These constants are proportional to positive powers of the
cosmological constant  $t$ and the powers are determined as follows.
The scaling dimension of the disk amplitude $< \hat{w}_0(l)>_0$ can be
estimated by KPZ-DDK argument \cite{kpz} \cite{dkd} to be $L^{-7/3}$, where
$L$ stands for the dimension of the boundary length.  $<\hat{w}_n(l_1, \cdots,
l_n)>_0$ can be obtained from $<\hat{w}_0(l)>_0$ by insertion of the
operator ${\cal H}(\sigma)$, which has dimension $L^{-4/3}$, at {\it n}
distinct points of the boundary.\cite{iikmns} Hence $D_n(l_1, \cdots,
l_n)$ and $T_n(l_1, \cdots, l_n)$ have dimension $L^{-(7 + 4n)/3}$.
Because $t$ has dimension $L^{-2}$, we find $a$ and $b$ are proportional to
$t^{5/6}$ and $t^{1/3}$, respectively.
Higher derivative terms are not included because the coefficients will
become  negative powers of $t$.  It is non-trivial but straightforward to show
$\tilde{T}_n(l_1,\cdots,l_n)$ also satisfies (\ref{9}) $\sim$ (\ref{11}).

Process 5 corresponds to the tadpole terms in the view point of
string field theory.\cite{iikmns}
Such terms contain the procuct of delta functions $\delta(l_j)$ and
their derivatives multiplied by a functional of the source functions $J_n$.
\cite{ike}
It  is rather difficult to determine those terms solely by integrability
conditions and we will not pursue this problem here.

We now turn to string field theory.  The continuum version of the matrix
model SDE is closely related to the string field Hamiltonian
and general form of the Hamiltonian in $c < 1$ string field theory
was presented in \cite{iikmns}.
The expression of the Hamiltonian was, however, rather formal because the
meaning of the summation over CFT states $|v>$ was not completely specified.
Here we will write down the string field Hamiltonian more explicitly.
For detailed discussion of the string
field theory in the temporal gauge we refer the reader to \cite{iikmns}
\cite{ik} \cite{ik2} \cite{ishi}.

General prescription is to express the generating functional in the form
\beqa
Z[J] & = & \lim_{D \rightarrow \infty} <0| e^{-DH} e^{S_{source}} |0>,
\label{13} \\
S_{source} & = & \int_0^{\infty} dl J_0(l) \Psi_0^{\dag}(l) +
   \sum_{n = 1}^{\infty} \int_0^{\infty} dl_1 \cdots \int_0^{\infty} dl_n
     J_n(l_1, \cdots,l_n) \Psi_n^{\dag}(l_1, \cdots,l_n).
\label{14}
\eeqa
Here $H$ is the string field Hamiltonian and $D$ is the geodesic distance
from the boundaries.  $\Psi_0^{\dag}(l)$ and $\Psi_n^{\dag}(l_1, \cdots,l_n)$
are the creation operators of the loops $\hat{w}_0(l)$ and
$\hat{w}_n(l_1,\cdots, l_n)$, respectively, and these satisfy the usual
commutation
relations with the corresponding annihilation operators
$\Psi_0$ and $\Psi_n$.
\beqa
\mbox{ [ } \Psi_0(l), \Psi_0 ^{\dag}(l') \mbox{ ] } & = &  \delta (l - l'), \n
\mbox{ [ } \Psi_n (l_1, \cdots, l_n),
             \Psi_n ^{\dag} ( l_1', \cdots, l_n ' ) \mbox{ ] }
 &  =  & \frac{1}{n} \{ \delta (l_1 - l_1') \cdots \delta (l_n - l_n')  \n
 & &       \quad \quad   +  \mbox{cyclic permutations} \}.
\label{15}
\eeqa
The vacuum state $|0>$ is defined by
\beq
\Psi_0(l) |0> = \Psi_n(l_1, \cdots, l_n) |0> = 0.
\label{16}
\eeq
Then the string field SDE can be obtained as the condition for the existence
of the limit (\ref{13}) and reads
\beq
 - \lim_{D \rightarrow \infty} \frac{\dee}{\dee D}
                              <0| e^{-DH} e^{S_{source}} |0>
 =  \lim_{D \rightarrow \infty} <0| e^{-DH} H e^{S_{source}} |0> = 0.
\label{17}
\eeq
Now $\Psi_n$ and $\Psi_n^{\dag}$ in $H$ can be replaced by $J_n$ and $D_n$,
respectively, and (\ref{17}) can be rewritten as  a differential equation
for $Z[J]$.
\beq
\hat{H} Z[J] = 0
\label{18}
\eeq
Here $\hat{H}$ is a differential operator with respect to $J$'s.
The string field SDE (\ref{18}) and the matrix model SDE will be equivalent
under suitable boundary conditions \cite{ik} if $\hat{H}$ is chosen as follows.
\beqa
\hat{H} & = & \int_0^{\infty} dl  J_0(l)l \tilde{T}_1(l) \n
& & + \sum_{n=1}^{\infty} n \int_0^{\infty} dl_1 \cdots \int_0^{\infty} d
l_{n+1} J_n(l_1+l_{n+1},l_2,\cdots,l_n) \tilde{T}_{n+1}(l_1,\cdots,l_{n+1}).
\label{19}
\eeqa
This yields the following string field Hamiltonian.
\beq
H = H_1 + H_2 + H_3 +H_4 + H_5, \label{20}
\eeq
\beqa
H_1 & = & \int_0^{\infty} dl_1 \int_0^{\infty} dl_2 \Psi_0^{\dag}(l_1) \Psi_0^{
\dag}(l_2) \Psi_0(l_1+l_2) (l_1+l_2) \n
& & + 2 \sum_{n=1}^{\infty} n \int_0^{\infty} dl \int_0^{\infty} dl_1 \cdots
\int_0^{\infty} dl_n \Psi_0^{\dag}(l) \Psi_n^{\dag}(l_1,\cdots,l_n) \Psi_n
(l_1+l,l_2,\cdots,l_n)l_1 \n
& & + \sum_{n=1}^{\infty} \sum_{m=1}^{\infty} (n+m) \int_0^{\infty} dl
\int_0^{\infty} dl' \int_0^{
\infty} dl_1 \cdots \int_0^{\infty} dl_n \int_0^{\infty} dl_1' \cdots \int_0^{
\infty} dl_m' \n
& & \Psi_n^{\dag}(l_1+l,l_2,\cdots,\l_n) \Psi_m^{\dag}(l_1'+l',l_2',\cdots,
l_m') \Psi_{n+m}(l_1+l',l_2,\cdots,l_n,l_1'+l,l_2'\cdots,l_m'), \nn
\eeqa
\beqa
H_2 & =& g \int_0^{\infty} dl_1 \int_0^{\infty} dl_2 \Psi_0^{\dag}(l_1+l_2)
\Psi_0(l_1) \Psi_0(l_2) l_1 l_2 \n
& & + 2g \sum_{n=1}^{\infty} n \int_0^{\infty} dl \int_0^{\infty} dl_1 \cdots
\int_0^{\infty} dl_n \Psi_n^{\dag}(l_1+l,l_2,\cdots,l_n) \Psi_n(l_1,\cdots,
l_n) \Psi_0(l) l l_1 \n
& & + g \sum_{n=1}^{\infty} \sum_{m=1}^{\infty} nm \int_0^{\infty} dl \int_0^{
\infty} dl' \int_0^{\infty} dl_1 \cdots \int_0^{\infty} dl_n \int_0^{\infty}
dl_1' \cdots \int_0^{\infty} dl_m' \n
& & \Psi_{n+m}^{\dag}(l_1+l,l_2,\cdots,l_n,l_1'+l',l_2',\cdots,l_m')
\Psi_n(l_1+l',l_2,\cdots,l_n) \Psi_m(l_1'+l,l_2',\cdots,l_m'), \nn
\eeqa
\beqa
H_3 & = & \int_0^{\infty} dl \Psi_1^{\dag}(l) \Psi_0(l) l \n
       & & + \sum_{n=1}^{\infty} n \int_0^{\infty} dl_1 \cdots \int_0^{\infty}
         dl_{n+1} \Psi_{n+1}^{\dag} (l_1,\cdots,l_{n+1})
             \Psi_n(l_1+l_{n+1},l_2,\cdots,l_n), \nn
\eeqa
\beqa
H_4 & =& 2a \mbox{ [ } \int_0^{\infty} dl \Psi_0^{\dag}(l) \Psi_1(l) \n
   & & + \sum_{n=2}^{\infty} n \int_0^{\infty} dl_1 \cdots \int_0^{\infty} dl_n
   \Psi_{n-1}^{\dag}(l_n+l_1,l_2,\cdots,l_{n-1}) \Psi_n(l_1,\cdots,l_n)
   \mbox{ ] } \n
   & & + b  \mbox{ [ }2 \int_0^{\infty} dl \{ \dee_l \Psi_0^{\dag}(l) \cdot
   \Psi_1(l) - \Psi_0^{\dag}(l) \cdot \dee_l \Psi_1(l) \} \n
  & & + \sum_{n=2}^{\infty} n \int_0^{\infty} dl_1 \cdots \int_0^{\infty} dl_n
   \{ (\dee_{l_1} + \dee_{l_n}) \Psi_{n-1}^{\dag}(l_n+l_1,l_2,\cdots,l_{n-1})
   \cdot \Psi_n(l_1,\cdots,l_n) \n
   & & - \Psi_{n-1}^{\dag}(l_n+l_1,l_2,\cdots,l_{n-1}) \cdot (\dee_{l_1}+
   \dee_{l_n}) \Psi_n(l_1,\cdots,l_n) \} \mbox{ ] }, \nn
\eeqa
\beqa
H_5 & = & \int_0^{\infty} dl  \rho_0(l) \Psi_0(l) \n
    & & + \sum_{n=1}^{\infty}  \int_0^{\infty} dl_1 \cdots \int_0^{\infty}
    dl_n \rho_n (l_1, \cdots,l_n) \Psi_n(l_1,\cdots,l_n).
\label{21}
\eeqa
Here we introduced the string coupling constant $g$, which has dimension
$L^{-14/3}$. The dimensions of fields are
$[ \Psi_n^{\dag} ] = L^{-(7+4n)/3} \quad (n \geq 0)$,
$[ \Psi_0 ] = L^{4/3}$ and $ [ \Psi_n ] = L^{(7+n)/3} \quad (n \geq 1)$.
The tadpole terms $H_5$ are also included in $H$ although the explicit form
of $\rho_n$ is left undetermined.  It is clear $H_n$ describes process n.

The string states $\Psi_0^{\dag}(l)|0>$ and $\Psi_n^{\dag}(l_1,\cdots,l_n)|0>$
which appear in the above string field theory do not cover the whole space
of string states. There exist more general mixed spin configurations
\beq
\frac{1}{N} tr(A^{m_1} B^{n_1} A^{m_2} B^{n_2} \cdots A^{m_s} B^{n_s})
\quad \quad (m_i, n_i \geq 1).
\label{22}
\eeq
Nonetheless SDEs (\ref{4}) satisfy the integrability condition and
the Hamiltonian (\ref{20}) defines a consistent string field theory. It
should also be possible to write down the continuum SDE for the loops
(\ref{22}) by a straightforward extension of the  present work and
to obtain the string field theory for the whole string space.  In a sense
we successfully truncated the string space.   As a result of this truncation
the world sheet in our string field theory does not contain two-dimensional
domains of down spins. Down spins are confined into one-dimensional regions,
trajectories of the locations of the inserted operator ${\cal H}$.
This is in sharp contrast to the world sheet in another formulation of the
c= 1/2 string field theory in \cite{ik2}.

It may be important to stress the necessity of the new process
4.  The string field Hamiltonian describes the time evolution of loops on the
world sheet.  Suppose sweeping the world sheet by loops of constant time.
When the loop of constant time meets a one-dimensional domain of down spins
along the way, the operator ${\cal H}$ is inserted (process 3).  This operator
will be present for a while, but eventually it will disappear at the other
end of the one-dimensional domain.  This is the process 4.  Hence it is quite
plausible to incorporate such a process into the Hamiltonian.  We have
shown this is indeed possible to do without changing the algebra (\ref{9})
$\sim$ (\ref{11}).

A remarkable fact is that the Hamiltonian (\ref{20}) is universal
for all c = 1 $-$ 6/(m(m+1)) strings.  c = 1 $-$ 6/(m(m+1)) string can
be realized by the (m $-$ 1)-matrix chain model.  The matrices $M_i$
are labeled by an integer $i (=1,\cdots,m-1)$ and the action has the form
\beq
S(M_1,\cdots,M_{m-1}) = \frac{1}{N} tr (\sum_{i=1}^{m-1} V_i(M_i) - c
\sum_{i=1}^{m-2} M_i M_{i+1}).
\label{23}
\eeq
It is obvious that if we consider correlation functions of the operator
\beq
\frac{1}{N} tr(M_1^{m_1} M_2 M_1^{m_2} M_2 \cdots M_1^{m_n} M_2),
\label{24}
\eeq
we obtain the same  SDEs as
(\ref{4}).  Hence the same string field Hamiltonian (\ref{20}).  The constants
 $a$ and $b$ are now given by
\beq
a = a_0 t^{1/m + 1/2} \mbox{  and  } b = b_0 t^{1/m},
\label{25}
\eeq
where $a_0$ and $b_0$ do not depend on the cosmological constant $t$.
Because the string field Hamiltonian is the same for any c, it might be
possible to go beyond the c=1 barrier by investigations along this line.

The constants $a_0$ and $b_0$ in (\ref{25}) remain unknown.
The tadpole terms $\rho_n$ are undetermined, either.  To determine these we
need to compute various amplitudes of $\hat{w}_n$. Necessity of relative
angle integrations in matrix chain models hinder the calculation of loop
amplitudes with mixed spin configurations on the loop boundaries and these
problems are left to the future investigations.

To recapitulate, we demonstrated that continuum SDEs (\ref{4}) are integrable
and they generate a closed algebra (\ref{9})$\sim$(\ref{11}).  Apparently
this algebra does not seem to be related to $W_{\infty}$ but it will
certainly be important to pursue its connection to $W_{\infty}$ further.
We also derived string field Hamiltonian from SDEs(\ref{4}) and this
Hamiltonian was found to be universal for all $c = 1 - 6/(m(m+1))$ strings.
Because there are already two different choices of the time
coordinate\cite{ik2}\cite{iikmns} in constructing $c < 1$ string field theory,
there may exist another choice in terms of which SDEs and string field
Hamiltonian have a manifest $W_{\infty}$ structure.

\vspace{1cm}
R.~N. thanks N.~Ishibashi and  H.~Kawai for discussions.

\newpage

{\large {\bf Figure Captions}}

\begin{description}
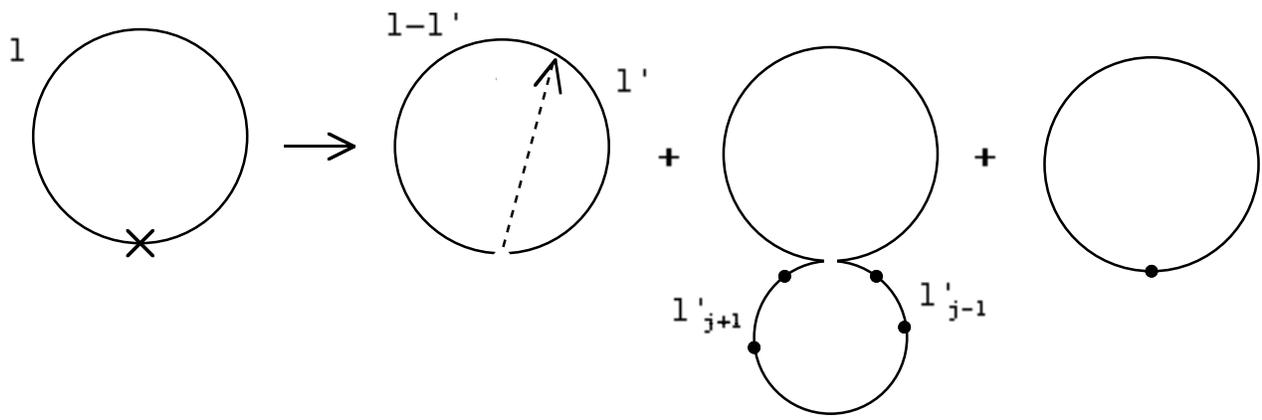
\item[Fig.1] Processes 1 $\sim$ 3 involved in SDE $T_1(l) Z[J] \approx 0$.
The cross on the left hand side represents the position of deformation.
The solid curve stands for the portion of the loop on which the spins
are up and the dot for insertion of ${\cal H}$ which changes the
configuration of spins.

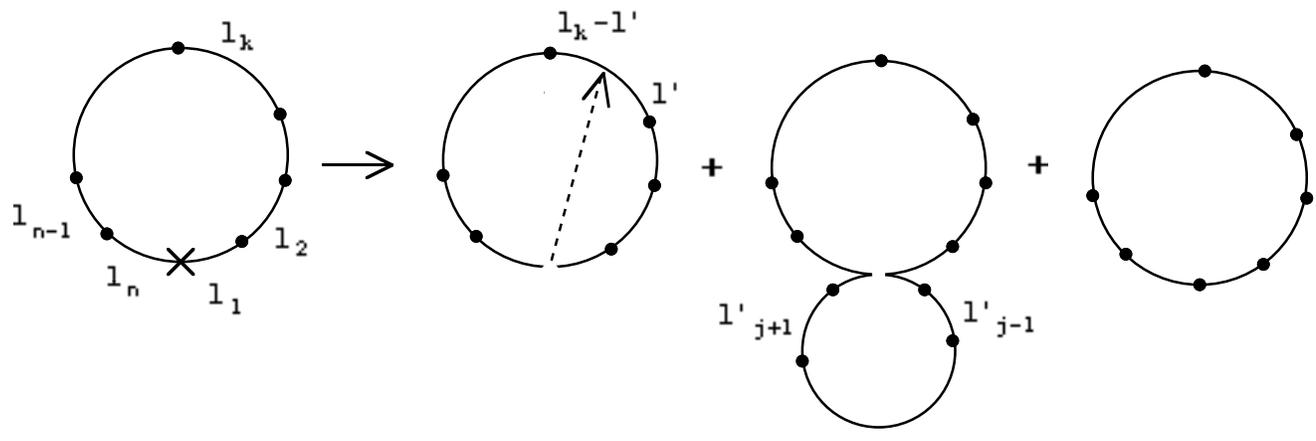
\item[Fig.2] Processes 1 $\sim$ 3 involved in SDE $T_n(l_1,\cdots,l_n) Z[J]
\approx 0$ $(n \geq 2)$.

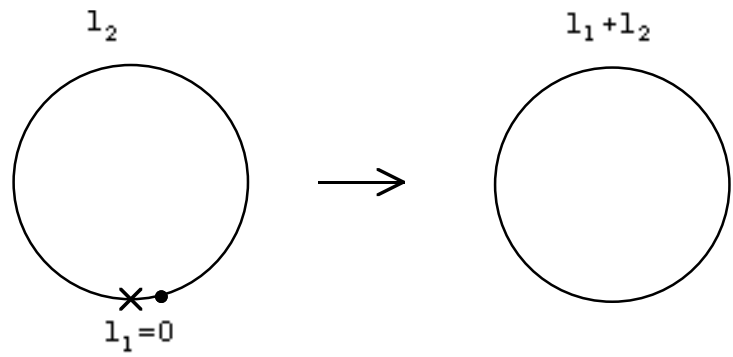
\item[Fig.3] Process 4 to be added to $T_2(l_1,l_2) Z[J]$.
If the position of deformation coincides with the position of ${\cal H}$,
{\it i.e.}, $l_1 = 0$, ${\cal H}$ may be removed.

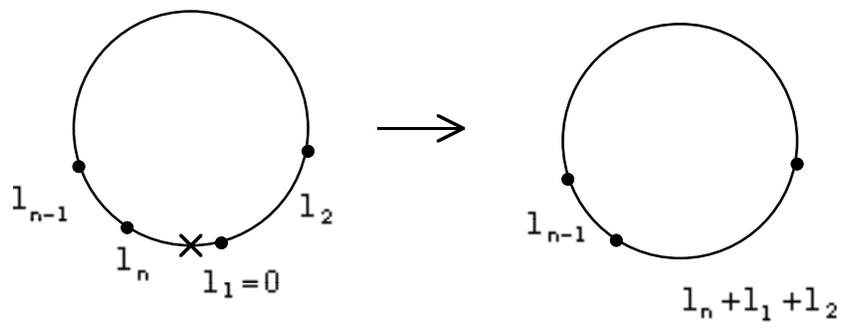
\item[Fig.4] Process 4 to be added to $T_n(l_1,\cdots,l_n) Z[J]$ $(n \geq 3)$.

\end{description}

\end{document}